\documentclass[prl,aps,twocolumn,superscriptaddress]{revtex4}
\usepackage{graphicx,amssymb,amsmath,color,psfrag}
\usepackage{amsthm}
\usepackage{amsfonts}
\usepackage{algorithmic}
\usepackage{enumerate}
\usepackage{latexsym}

\begin{document}

\title{Quantum Monte Carlo study of a dominant $s$-wave pairing symmetry in
    iron-based superconductors}
%\author{Tianxing Ma,$^{1,2}$ Hai-Qing Lin$^{2}$ and Jiang-Ping Hu$^{3,4}$}
\author{Tianxing Ma}
\affiliation{Department of Physics, Beijing Normal University,
Beijing 100875, China\\}

\affiliation{Beijing Computational Science Research Center,
Beijing 100084, China}

\author{Hai-Qing Lin}
\affiliation{Beijing Computational Science Research Center,
Beijing 100084, China}

\author{Jiangping Hu}
\email{jphu@iphy.ac.cn}
\affiliation{Beijing National Laboratory for Condensed Matter Physics, Institute of
Physics, Chinese Academy of Sciences, Beijing 100080, China}
\affiliation{Department of Physics, Purdue University, West Lafayette, Indiana
47907, USA }
\date{\today}

%\affiliation{$^{1}$Department of Physics, Beijing Normal University, Beijing 100875, China \\
%$^{2}$Beijing Computational Science Research Center, Beijing 100084, China\\
%$^{3}$Beijing National Laboratory for Condensed Matter Physics, Institute of
%Physics, Chinese Academy of Sciences, Beijing 100080, China \\
%$^{4}$Department of Physics, Purdue University, West Lafayette, Indiana
%47907, USA \\}

\begin{abstract}
{\  We perform a systematic quantum Monte Carlo study of the
pairing correlation in the $S_4$ symmetric microscopic model for iron-based superconductors.  It is found that the pairing with an extensive $s$-wave  symmetry robustly dominates over other pairings at low temperature in reasonable parameter region regardless of the change of Fermi surface topologies. The pairing susceptibility, the effective pairing
interaction and the $(\pi,0)$
antiferromagnetic(AFM) correlation  strongly increase as the on-site Coulomb
interaction increases, indicating the importance of the effect of electron-electron correlation.  Our non-biased
numerical results provide a unified understanding of superconducting
mechanism in iron-pnictides and iron-chalcogenides and demonstrate that the superconductivity is driven by strong electron-electron correlation effects. }
\end{abstract}

\maketitle

%\pacs{73.61.Wp, 73.20.At, 73.21.-b }
A today's major challenge in the study of iron-based superconductors\cite{Hosono,ChenXH,nlwang,ChenXL} is how to obtain an unified microscopic understanding of the different families of these materials, in particular, iron-pnicitides and iron-chalcogenides,  which distinguish themselves from each other with distinct Fermi surface topologies\cite{Wang_122Se, Zhang_122Se,Mou_122Se}.   In the past several years, the majority of the theoretical studies of iron-based high temperature superconductors were based on models with complicated multi-d orbital band structures\cite{john,hirschfeld,Dongj2008,Mazin2008,Kuroki2011, WangF, thomale1, thomale2, chubukov,zlako,Arita2009}.  The conclusions  from these studies provided a good understanding of iron-pnictides while  drawing a very different picture regarding of the magnetism and superconductivity of iron-chalcogenides because of their strong dependence on  theoretical  approximations and the topology of Fermi surfaces. Effective models emphasizing local AFM exchange couplings appear to unify the understanding of superconducting states of both materials\cite{Maiti2011,seo2008,Si2008,Fang2008nematic,Ma2008lu,hu1,hu4,luxl,berg}.
However, they lack of a support from  more fundamental  microscopic electronic physics, and using some unbiased numerical techniques is believed to be the only opportunity to win this great challenge as Hartree-Fork type approaches are biased
if the electronic correlation dominates in the system.
%f the system is a strongly correlation electron system.
 %and highly sensitive to the change of Fermi surfaces.

Recently, it has been shown that the underlining electronic structure in iron-based superconductors,
which is responsible for superconductivity at low energy, is essentially governed by a two-orbital model with a $S_4$ symmetry, the symmetry of  the building block-the trilayer FeAs or FeSe structure  in iron-based superconductors. In the model, the dynamics
of two $S_4$ iso-spin components are weakly coupled so that the essential physics is controlled by  a single $S_4$ iso-spin component. Thus,  a minimum effective model that captures the low energy electronic and magnetic properties  can be well described by an  extended one-orbital Hubbard model near half-filling\cite{Hu2012s4}. Such a microscopic understanding  provides a new opportunity to make  use of highly controllable and unbiased numerical methods to study iron-based superconductors, in particular, to  obtain a possible unified understanding of iron-pnicitides and iron-chalcogenides.
\begin{figure}[tbp]
\includegraphics[scale=0.375]{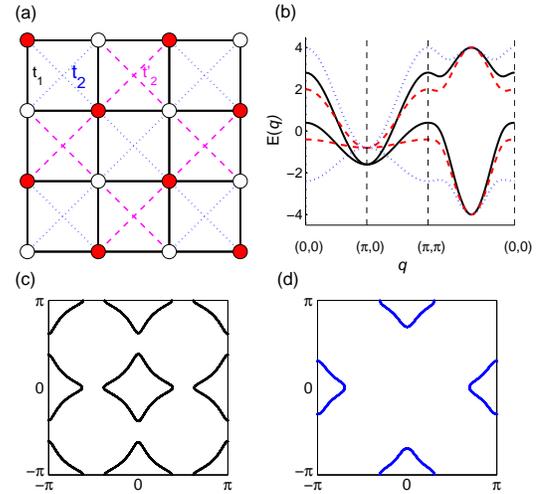}
\caption{(Color online)(a) Here red and white circles indicate
different sub-lattice A and B. The dark solid lines indicate $t_1$, dot blue lines
indicates $t_2$, and dash purple lines indicates $t^{\prime }_2$.
(b) The energy band along high symmetry line in unfolded Brillouin Zone. Solid dark line: $t_1$=0.3, $t_2$=1.4, $t'_2$=-0.6; dash red line:  $t_1$=0.3, $t_2$=1.2, $t'_2$=-0.8 and
dot blue line: $t_1$=0.8, $t_2$=1.2, $t'_2$=-0.8\cite{t1}.
(c) Fermi surface at half filling for $t_{1}=0.3, t_{2}=1.4, t'_2=-0.6$ (a typical case for iron-pnictides\cite{hding,hding2,lxyang,ludh,chenf}), and (d) at electron filling $<n>=1.1$ for $t_{1}=0.8, t_{2}=1.2, t'_2=-0.8$( a typical case for iron-chalcogenides\cite{Wang_122Se, Zhang_122Se,Mou_122Se}).}
\label{Fig:Sketch}
\end{figure}

In this Letter, we perform a systematic quantum Monte Carlo study of the pairing correlation in the $S_4$ symmetric microscopic model. We find  that the pairing with an extensive $A_{1g}$ $s$-wave
 symmetry robustly dominates over other pairings at low temperature in reasonable parameter region regardless of the change of Fermi surface topologies, the presence or absence of hole pockets at $\Gamma$ point.  For both iron-pnictides and iron-chalcogenides, the pairing susceptibility, the effective pairing
interaction and the $(\pi,0)$ AFM correlation  strongly increase as the on-site Coulomb
interaction increases. Our study demonstrates that the superconductivity in iron-based superconductors is driven by  electron-electron correlation and  the nesting between electron and hole pockets  is not an essential physics in iron-based superconductors.  This conclusion is behind many proposed effective models\cite{Maiti2011,seo2008,Si2008} but   could not be conclusively reached  with approximated methods\cite{Arita2009}.  The fact that  the extended $s$-wave is favored even in the case  without hole pockets  differs from  the earlier simple conjecture in \cite{Mazin2008,Kuroki2011}.  Our unbiased numerical results thus present a rather different picture form mean field approaches.
%, and advance the entire iron-based superconductors.

As shown in Fig. \ref{Fig:Sketch} (a), the minimum extended Hubbard model for a single $S_4$ iso-spin component in the iron-square lattice  is  described by
\begin{eqnarray}
&&H=t_{1}\sum_{\mathbf{i}\eta \sigma }(a_{\mathbf{i}\sigma }^{\dag }b_{%
\mathbf{i}+\eta \sigma }+h.c.) \notag \\
&&+ t_{2}[\sum_{\mathbf{i}\sigma }a_{\mathbf{i}\sigma }^{\dag }a_{\mathbf{i}%
\pm (\hat{x}+\hat{y}),\sigma }+\sum_{\mathbf{i}\sigma }b_{\mathbf{i}\sigma }^{\dag }b_{\mathbf{i}\pm (\hat{x}-\hat{y})\sigma
}] \notag \\
&&+t_{2}^{\prime }[\sum_{\mathbf{i}\sigma }a_{\mathbf{i}\sigma }^{\dag }a_{%
\mathbf{i\pm }(\hat{x}-\hat{y})\sigma }+\sum_{\mathbf{i}\sigma }b_{\mathbf{i}\sigma }^{\dag }b_{\mathbf{i\pm (}%
\hat{x}+\hat{y})\sigma }]\notag\\
&& +U\sum_{i}(n_{ai\uparrow}n_{ai\downarrow} + n_{bi\uparrow}n_{bi\downarrow})
+\mu\sum_{i\sigma}(n_{ai\sigma }+n_{bi\sigma })
\end{eqnarray}
Here, $a_{i\sigma}$ ($a_{i\sigma}^{\dag}$) annihilates (creates) electrons
at site $\mathbf{R}_i$ with spin $\sigma$ ($\sigma$=$\uparrow,\downarrow$)
on sublattice A, $b_{i\sigma}$ ($b_{i\sigma}^{\dag}$) annihilates (creates)
electrons at the site $\mathbf{R}_i$ with spin $\sigma$
($\sigma$=$\uparrow,\downarrow$) on sublattice B,
$n_{ai\sigma}=a_{i\sigma}^{\dagger}a_{i\sigma}$,
$n_{bi\sigma}=b_{i\sigma}^{\dagger}b_{i\sigma}$, $\eta=(\pm \hat{x},0)$ and $(0,\pm \hat{y})$.
%$t_1$ is the NN hopping integral and $\mu$ the chemical potential.
In  the above model, for simplicity and clarity, we only keep a minimum set of parameters which include three key shortest hopping parameters that are responsible for the physical picture revealed by the $S_4$ symmetry\cite{Hu2012s4}.
The selection of parameters in following studies does capture the essential physics of typical cases for iron-pnictides\cite{hding,hding2,lxyang,ludh,chenf} and iron-chalcogenides\cite{Wang_122Se, Zhang_122Se,Mou_122Se}, as shown in Fig.1 (b-d).
\begin{figure}[tbp]
\includegraphics[scale=0.425]{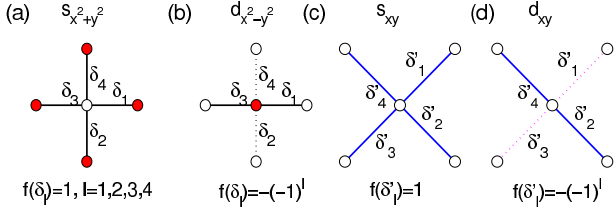}
\caption{(Color online) Phase of the $s_{x^2+y^2}$, $d_{x^2-y^2}$, $s_{xy}$ and $d_{xy}$. }
\label{Fig:Pairing}
\end{figure}

Our numerical calculations were mainly performed on an $8^2$ or a $12^2$ lattice  with periodic boundary conditions. The system was simulated using determinant quantum Monte Carlo (DQMC) at finite temperature.
The basic strategy of DQMC is to express
the partition function as a high-dimensional integral over a set of
random auxiliary fields. The integral is then accomplished by
Monte Carlo techniques. In
our simulations, 8000 sweeps were used to equilibrate the system. An
additional 45000 sweeps were then made, each of which generated a
measurement. These measurements were split into fifteen bins which provide the
basis of coarse-grain averages and errors were estimated based on standard
deviations from the average. For more technique details we refer to
Refs.~\cite{Blankenbecler1981,Maqmc,Maqmc2}.

As magnetic excitation might play an important role in the superconducting (SC) mechanism
of electronic correlated systems, we first studied the magnetic
correlations in such system. We define the spin susceptibility in the
$z$ direction at zero frequency,
\begin{eqnarray}
\chi(q) = \int_{0}^{\beta}d\tau \sum_{d,d'=a,b} \sum_{i,j}
e^{iq\cdot(i_{d}-j_{d'})} \langle\textrm{m}_{i_{d}}(\tau) \cdot
\textrm{m}_{j_{d'}}(0)\rangle
\end{eqnarray}
where $m_{i_{a}}(\tau)$=$e^{H\tau} m_{i_{a}}(0) e^{-H\tau}$ with
$m_{i_{a}}$=$a^{\dag}_{i\uparrow}a_{i\uparrow}-a^{\dag}_{i\downarrow}a_{i\downarrow}$
and
$m_{i_{b}}$=$b^{\dag}_{i\uparrow}b_{i\uparrow}-b^{\dag}_{i\downarrow}b_{i\downarrow}$.
%We measure $\chi$ in unit of $\mid$$t$$\mid$$^{-1}$.

To investigate the SC property of iron-based superconductors, we computed the pairing
susceptibility,
\begin{equation}
P_{\alpha }=\frac{1}{N_{s}}\sum_{i,j}\int_{0}^{\beta }d\tau \langle \Delta
_{\alpha }^{\dagger }(i,\tau )\Delta_{\alpha }^{\phantom{\dagger}%
}(j,0)\rangle ,
\end{equation}
where $\alpha $ stands for the pairing symmetry. Due to the constraint of
on-site Hubbard interaction in Eq.~(1), pairing between two sublattices is
favored and the corresponding order parameter $\Delta_{\alpha }^{\dagger
}(i)$\ is defined as
\begin{equation*}
\Delta _{\alpha }^{\dagger }(i)\ =\sum_{l}f_{\alpha }^{\dagger }(\delta_{l})(a_{{i}\uparrow }b_{{i+{\delta_{l}}}\downarrow }-a_{{i}\downarrow }b_{{%
i+\delta_{l}}\uparrow })^{\dagger },
\end{equation*}%
with $f_{\alpha}(\mathbf{{\delta_l})}$ being the form factor of pairing
function. Here, the vectors $\mathbf{{\delta_l}}$ ($l$=1,2,3,4) denote
the nearest neighbour (NN) inter sublattice connections where $\mathbf{\delta}$ is $(\pm \hat{x},0)$ and $(0,\pm \hat{y})$,  or the next nearest neighbour (NNN) inner sublattice connections
where $\mathbf{\delta'}$ is $\pm (\hat{x},\hat{y})$ and $\pm(\hat{x},-\hat{y})$.

%Due to the structure of square lattice,
%there are two types of pairing form.
We study four kinds of pairing form, as that sketched in Fig. \ref{Fig:Pairing}. For $s_{x^2+y^2}$-wave pairing, $f_{s}(%
\mathbf{\delta _{l}})=1$. For $d_{x^{2}-y^{2}}$ pairing, $f_{d}(\mathbf{%
\delta _{l}})$ is 1 when $\mathbf{\delta _{l}}=(\pm \hat{x},0)$ and -1 otherwise.

Another two interesting pairing forms are $d_{xy}$-wave and extensive $s_{xy}$-wave,
%&\text{:}
\begin{eqnarray}
&\text{$d_{xy}$-wave}&: f_{d_{xy}}(\mathbf{\delta'_{l}})=1  ( \mathbf{\delta'_{l}}=\pm(\hat{x},\hat{y})) \notag \\
&\text{and}& f_{d_{xy}}(\mathbf{\delta'_{l}})=-1  (\mathbf{\delta'_{l}}=\pm(\hat{x},-\hat{y})),  \notag \\
&\text{$s_{xy}$-wave}&: f_{s_{xy}}(\mathbf{\delta'_{l}})=1,~l=1,2,3,4.
\end{eqnarray}

\begin{figure}[tbp]
\includegraphics[scale=0.425]{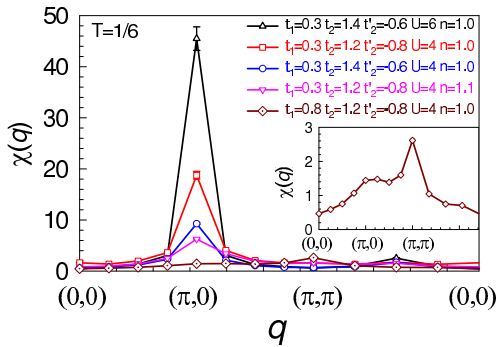}
\caption{(Color online) Magnetic susceptibility $\chi(q)$ versus
momentum $q$ on an $8^2$ lattice.}
\label{Fig:spin}
\end{figure}

In Fig. \ref{Fig:spin}, we present the spin susceptibility
$\chi(\textbf{q})$ at different electron fillings and $U$ for temperature $T$=1/6.
A sharp peak at $(\pi,0)$ in Fig.~\ref{Fig:spin} indicates the existence of
AFM spin correlation in iron-based superconductors close to half filling.
From Fig.~\ref{Fig:spin}, one can notice that at half-filling,
the peak is very sharp, and as the electron filling $<n>$
decreases from half filling, $\chi(\textbf{q})$ is reduced around
the $(\pi, 0)$ point, which indicates that
the AFM spin correlation is suppressed when the system is doped away from
half filling. The $(\pi,0)$ AFM spin fluctuations have been universally observed in all iron-based superconductors\cite{spinwave-dela2008,zhao1,zhaojun2012,lip2,my,wangmy2012,hu2012u}. It is clear that our model and the results naturally provides an explanation for the stable AFM exchange couplings $J_2$ observed by neutron scattering\cite{zhao1,lip2,my} in parental compounds.

From the behavior of magnetic correlation shown in Fig.~\ref{Fig:spin},
it is also important to note that the $(\pi,0)$ spin correlation does not depend on the presence of
the hole pockets at $\Gamma$ point, which indicates that such an AFM correlation is driven by electron-electron correlation rather than nesting between hole and electron Fermi pockets\cite{Arita2009}. %{\color{blue}
%as that demonstrated in Ref..
%}.
The  $(\pi,0)$ AFM correlation is also stabilized by the fact that the diagonal or NNN hopping parameters $t_2$ are larger than the NN hopping $t_1$. As shown in Fig.~\ref{Fig:spin}, if $t_1$ is significantly large, the $(\pi,\pi)$ magnetic correlation can be dominant. As we will show later, in this case, $d_{x^2-y^2}$ pairing can be significantly enhanced.
\begin{figure}[tbp]
\includegraphics[scale=0.375]{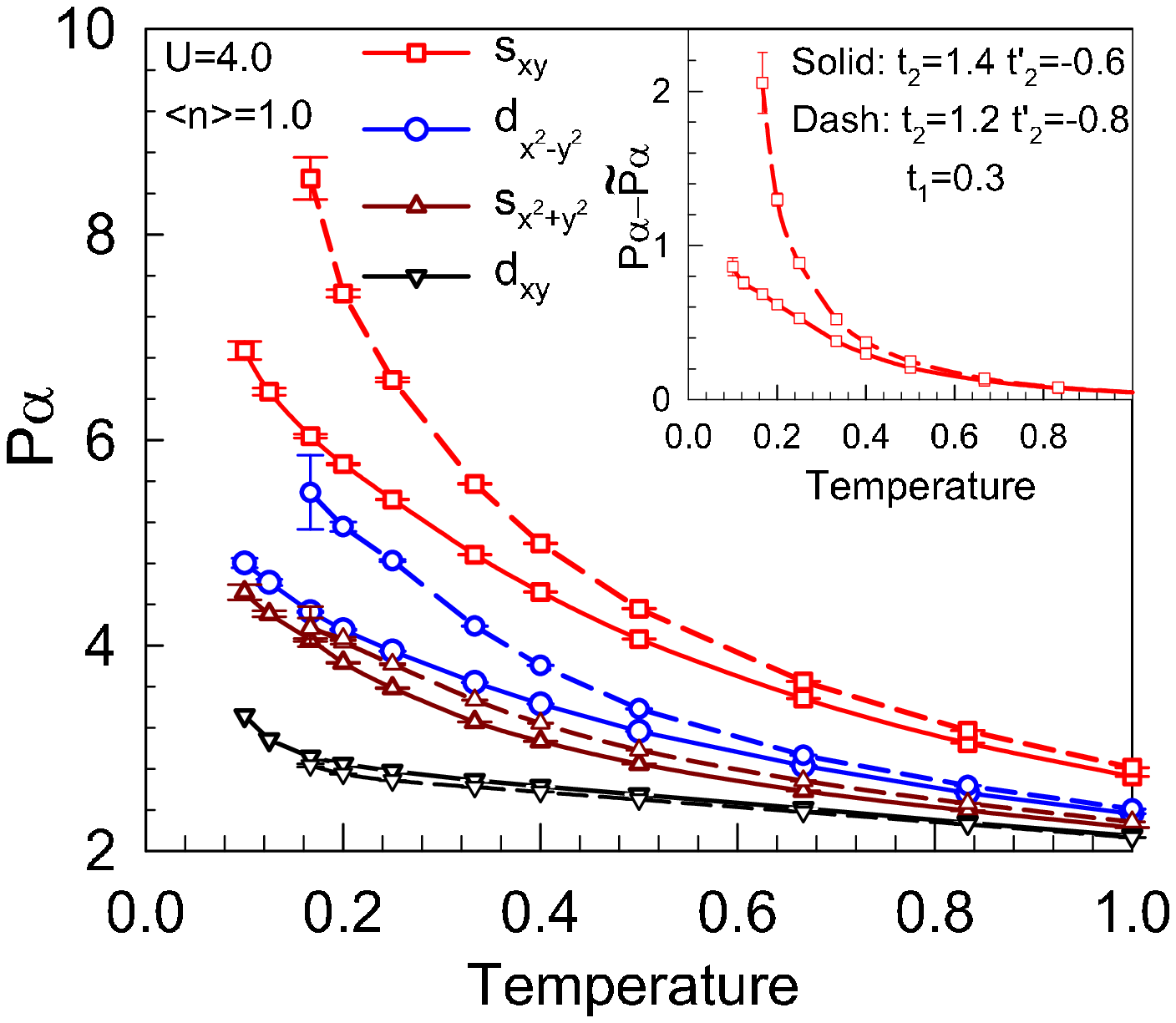}
\caption{(Color online) Pairing susceptibility $P_{\alpha}$ as a function
of temperature for different pairing symmetries at $t_1=0.3, t_2=1.4, t'_2=-0.6$ (solid line) and
$t_1=0.3, t_2=1.2, t'_2=-0.8$ (dash line) with $U=4.0$ and $<n>$=1.0 on an $8^2$ lattice.
In the inset: The effective pairing interaction $P_{s_{xy}}-\widetilde{P}_{s_{xy}}$.}
\label{Fig:FigP}
\end{figure}

Fig.~\ref{Fig:FigP} shows the temperature dependence of pairing
susceptibilities for different pairing symmetries.
 Within the parameter range
investigated, the pairing susceptibilities for various pairing
symmetries increase as the temperature is lowered. Most remarkably,
$s_{xy}$ increases much faster than any other pairings symmetry at low temperatures.
This demonstrates that the $s_{xy}$ pairing symmetry is dominant over the
other pairing symmetry near half filling.
%{\color{red}
When other parameters are fixed, reducing $t_{2s}=(t_2+t'_2)/2$ from 0.4 to 0.2, or increasing the absolute value of $t'_2$,
one may also see that pairing susceptibilities with different symmetries are all enhanced, in particular, the $s_{xy}$ pairing
susceptibility.

%}
%$S_xy$ and $d_xy$ Moreover, keep with the increasing of
%Fig.~\ref{Fig:Fig1} (b) again supports
%that the $s_{xy}$ should dominate.

In order to extract the effective pairing interaction in
different pairing channels, the bubble contribution $\widetilde{P}
_{\alpha }(i,j)$ is also evaluated, which is achieved by replacing $\langle
a_{{i}\downarrow }^{\dag }b_{{j}\uparrow }a_{i+\delta_{l}\downarrow}^{\dag}
b_{j+\delta_{l'}\uparrow}\rangle $ with $\langle a_{{i}\downarrow }^{\dag
}b_{{j}\uparrow }\rangle \langle a_{i+\delta_{l}\downarrow }^{\dag }
b_{j+\delta_{l'}\uparrow }\rangle $ in Eq.~(3).
In the inset of Fig. \ref{Fig:FigP}, we plot $P_{s_{xy}}-\widetilde{P}_{s_{xy}}$ for
$t_1=0.3, t_2=1.4, t'_2=-0.6$ (solid line). %and $t_1=0.3, t_2=1.2, t'_2=-0.8$ (dash line) .
It is apparent that  $P_{s_{xy}}-\widetilde{P}_{s_{xy}}$ shows a very similar temperature
dependence to that of $P_{s_{xy}}$.
The effective pairing interaction for $P_{s_{xy}}$,
is found to take a positive value and to increase with lowering temperature.
The positive effective pairing interaction indicates that there actually
exists attraction for the $s_{xy}$ pairing.
%{\color{red}
The effective $s_{xy}$ pairing interaction for $t_1=0.3, t_2=1.2, t'_2=-0.8$
is also shown in Fig.\ref{Fig:FigP} as dash line.
Comparing results with different $t'_2$, the effective $s_{xy}$ pairing interaction is enhanced
greatly as the absolute value of $t'_2$ increases.
\begin{figure}[tbp]
\includegraphics[scale=0.35]{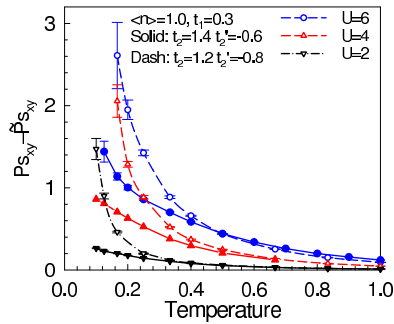}
\caption{(Color online) The effective pairing interaction
$P_{s_{xy}}-\tilde{P}_{s_{xy}}$ as a function of temperature for different $U$
at $t_1=0.3, t_2=1.4, t'_2=-0.6$ (solid line) and $t_1=0.3, t_2=1.2, t'_2=-0.8$ (dash line) on an $8^2$ lattice.}
\label{Fig:FigU}
\end{figure}

%\begin{figure}[tbp]
%\includegraphics[scale=0.5]{Fig5.eps}
%\caption{(Color online) The effective pairing interaction
%$P_{\alpha}-\tilde{P}_{\alpha}$ as a function of temperature for different pairing symmetries
%at $<n>$=1.0 for $U=2.0$, $U=4.0$ and $U=6.0$ with (a) $t_1=0.3, t_2=1.4, t'_2=-0.6$ and (b)  $t_1=0.3, t_2=1.2, t'_2=-0.8$.}
%\label{Fig:Fig3}
%\end{figure}

In Fig.~\ref{Fig:FigU}, we present the effective pairing interaction as a
function of temperature for $P_{s_{xy}}$ at different $U$ with
 $t_1=0.3, t_2=1.4, t'_2=-0.6$ (solid line) and (b) $t_1=0.3, t_2=1.2, t'_2=-0.8$ (dash line).
One can see that, the effect pairing interaction is enhanced as $U$ increases.
Especially, the effective $s_{xy}$ pairing interaction shown in Fig. \ref{Fig:FigU} tends to diverge in low temperatures,
and the increasing $U$ tends to promote such diverge.
This demonstrates that the  electron-electron correlation plays a key  role in
driving the superconductivity.
\begin{figure}[tbp]
\includegraphics[scale=0.475]{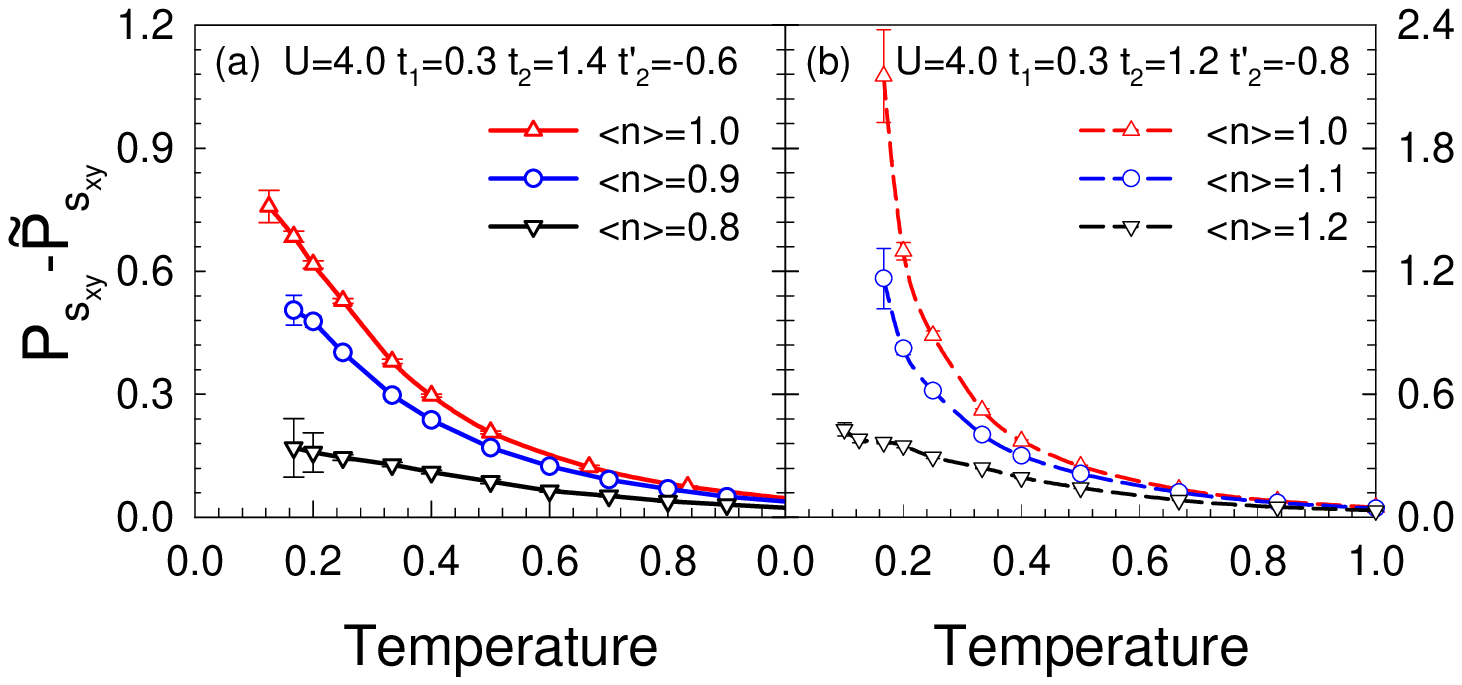}
\caption{(Color online) The effective paring interaction $P_{s_{xy}}-\tilde{P}_{s_{xy}}$ as a function of temperature for different electron fillings
at $t_1=0.3, t_2=1.4, t'_2=-0.6$ (solid lines) and $t_1=0.3, t_2=1.2, t'_2=-0.8$ (dash lines) on an $8^2$ lattice. %Inset: The pairing susceptibility.
}
\label{Fig:Fign}
\end{figure}

We also studied the temperature dependence of effective pairing interaction at different electron fillings.
In Fig. \ref{Fig:Fign}, the temperature dependence of effective pairing interaction is shown for
$<n>$=1.0, 0.9 and 0.8 with $t_1=0.3, t_2=1.4, t'_2=-0.6$ (a); and $<n>$=1.0, 1.1 and 1.2
with $t_1=0.3, t_2=1.2, t'_2=-0.8$(b). Both Fig. \ref{Fig:Fign} (a) and (b) show that, the effective paring interaction
decreases as the system is doped away from half filling.
As shown in Figs. (3,4,6), the decrease of the peak at ($\pi, 0$) of spin susceptibility is correlated with the suppression of the pairing susceptibility. This directly confirms that the $(\pi,0)$ AFM fluctuations favors the $s_{xy}$ pairing.
% as that shown in Fig. ~\ref{Fig:Fig4} (b).
%\begin{figure}[tbp]
%\includegraphics[scale=0.5]{Fig5.eps}
%\caption{(Color online) (a) The Pairing susceptibility
%$P_{\alpha}$ as a function of temperature for different pairing symmetries
%at $t_1=0.3, t_2=1.4, t'_2=-0.6$ and $<n>$=1.0, $<n>$=0.9, and $<n>$=0.8 with $U=4.0$.
%(b) The effective pairing interaction $P_{\alpha}-\tilde{P}_{\alpha}$ at different electron fillings. }
%\label{Fig:Fig4}
%\end{figure}
%Fig. ~\ref{Fig:Fig5} shows the temperature dependence of  pairing
%susceptibility (a) and effective pairing interaction (b) for $t_1=0.3, t_2=1.2, t'_2=-0.8$ at
% $<n>$=1.0, $<n>$=1.1, and $<n>$=1.2.

%
%

%{\color{red} Reminding the energy band shown in Fig. 1(b), the model unifies the iron-pnictides and iron-chalcogendies.
%When other parameters are fixed, increasing the absolute value of $t'_2$ can flatten the dispersion along $\Gamma$-$M$ direction of energy and cause the hole pocket completely vanishes, and the $d_{x^2-y^2}$ is expected to be stronger in iron chalcogenides than in iron-pnictides as the hole packet is suppressed. Our numerical results exactly confirm this prediction.

\begin{figure}[tbp]
\includegraphics[scale=0.35]{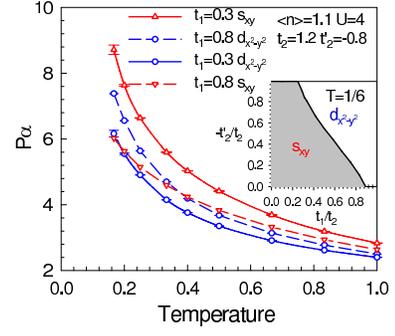}
\caption{(Color online) $P_{\protect\alpha}$ as a function of temperature
at $<n>$=1.1, $U=4.0$, $t_2=1.2$, $t'_2=-0.8$ for different $t_1$ on an $8^2$ lattice.
Inset: the competition between $s_{xy}$ and $d_{x^2-y^2}$ depends on $t_1/t_2$ and $-t'_2/t_2$.}
\label{Fig:Figt1}
\end{figure}

Following, we reported the effect of %the nearest neighbor hopping
$t_1$ on pairing symmetry. In Fig. \ref{Fig:Figt1}, the pairing behavior for different $t_1$ is shown for $<n>=1.1$.
As $t_1$ increases significantly,  it is possible that the $d_{x^2-y^2}$ wave pairing becomes dominant. A global phase diagram of the leading pairing symmetry  in the $t_1/t_2$-$|t_2'|/t_2$ plane, obtained by DQMC at temperature $T=1/6$,  is shown in the inset of Fig. \ref{Fig:Figt1}. This phase diagram indicates that the $s_{xy}$ is robust when $t_1$ is small. The phase transition line in the inset is roughly corresponding to the hopping parameter setting when the dispersion of the  band at $\Gamma$ which is originally responsible for the hole pocket gets reversed. Since such a reverse does not take place in the band structure of iron-based superconductors\cite{hding}, the $s_{xy}$-wave thus is robust.  The robustness is further shown in  Fig. \ref{Fig:Figt1} where we reported DQMC calculations throughout the $t_1/t_2
$ vs $t_2^{\prime }/t_2$ phase diagram.
%  For most results presented here, we selected relative large value of $t_2^{\prime }$ in order to show the robustness of the $s$-wave pairing and prove that it is insensitive to the change of Fermi
%surfaces.
In general,  the hopping parameters in the $S_4$ model is $t_2>t_1>|t_2^{\prime }|$, and the fitting to band structure shows that $t_1$ is roughly $%
0.4 t_2$ and $t_2^{\prime }$ is around $0.3t_2$\cite{Hu2012s4}.
%When $t_2^{\prime }$ is smaller, $s_{xy}$-wave is more stable as
%that shown in Fig 7.

%are reporting in another paper.}

\begin{figure}[tbp]
\includegraphics[scale=0.35]{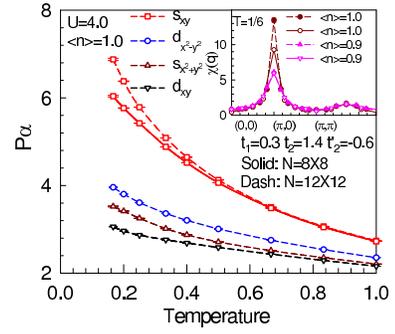}
\caption{(Color online) $P_{\protect\alpha}$ as a
function of temperature for different pairing symmetries at $t_1=0.3,
t_2=1.4, t^{\prime }_2=-0.6$ of a $12^2$ (dash line) and an $8^2$(solid line) lattice with $U=4.0$ and $<n>$=1.0. In the inset: $\protect\chi(q)$ versus
 $q$ for a $12^2$ (dash line with solid symbols ) and $8^2$ (solid line with open symbols) lattice at $<n>=1.0$ (dark red) and $%
<n>=0.9$ (pink).}
\label{Fig:Lattice}
\end{figure}

Finally, we present the
pairing susceptibility for different pairing symmetries on a $12^2$ lattice
and compare them to the $s_{xy}$ paring susceptibility on an $8^2$ lattice in Fig. \ref{Fig:Lattice}.
One can see that, the results for a larger lattice size confirm that the $%
s_{xy}$ dominates over other kinds of pairing symmetry. Moreover, it is
interesting to see that, the $s_{xy}$ pairing susceptibility increases as
the lattice size increases, especially at low temperature. This enhancement
is consistent with the behavior of spin susceptibility $\chi(\textbf{q})$ shown in the inset of Fig. \ref{Fig:Lattice}, in which $\chi(\textbf{q})$ is also enhanced as the lattice size increases. The peak of $\chi(\textbf{q})$  at $(\pi, 0)$ point increases as the lattice size increases at $<n>=1.0$, which implies there may be a static
magnetic order develops for the ground state near half filling.
%Away from
%half filling, the magnetic susceptibility is clearly weakened, which means that the static magnetic order is likely destroyed.
The above results of the pairing susceptibility have also shown to be valid when long range pairing correlation are calculated\cite{Macpmc}. Thus, the $8^2$ lattice is large enough to investigate the dominant pairing symmetry.

% Thus, the $8^2$ lattice is large enough to investigate the dominant pairing symmetry.  }

Overall, these results clearly suggest that the superconductivity and pairing symmetry in iron-based superconductors are determined by the combination of strong electron-electron correlation and the microscopic setting of hopping parameters. % {\color{blue}
In the model, if $t_2$ and $t_2'$ are fixed, a small decrease of $t_1$ can cause the vanishing of hole pockets at $\Gamma$. Since the $d$-wave pairing channel is caused by $t_1$, we can conclude that the $s$-wave pairing is ever more robust in iron-chalcogenides than in iron-pnictides,
a result completely different from weak coupling approaches. Recent experimental results by angle resolved photoemission spectroscopy have strongly suggested that pairing symmetry in both iron-pnictides\cite{hding} and iron-chalcogenides is a $s$-wave\cite{wangxp2012}. Our study clearly provides such a unified microscopic understanding.

In summary, we study the paring susceptibility and effective paring interaction in  iron-based superconductors
based on an effective $S_4$ model\cite{Hu2012s4}. It is confirmed that the $(\pi,0)$ AFM dominates at half filling and
the paring susceptibility with $s_{xy}$ symmetry are enhanced as the electron-electron correlation increases,
especially at low temperature.
It is suggested that the $s_{xy}$ wave paring symmetry, which is the $A_{1g}$ phase corresponding to the point group of the lattice and homogenous with respect to  the transitional  symmetry of the  iron lattice4\cite{pairing},  should be dominant in  iron-based superconductors. The reported strong AFM at half filling and the behavior of pairing susceptibility and effective pairing
interaction strongly support that  the microscopic superconducting mechanism for cuprates and iron-based superconductors including both iron-pnictides and iron-chalcogenides are identical.

{\it Acknowledgement:}  JP thanks H. Ding, D.L. Feng  for useful discussion.
The work is supported  by the Ministry of Science and Technology of China 973 program (2011CB922200 and 2012CB821400),  NSFCs (Grant. No. 1190024 and No. 11104014), and
Research Fund for the Doctoral Program of Higher Education of China 20110003120007.

%as well as the increasing
%of

%In the whole temperature
%regime, one can observe that the value of $P_{d+id}$ at $U=3|t|$ is smaller
%than the corresponding noninteracting one ($U=0$), as displayed in
%Fig.~\ref{Fig:DS}(a), which reflects the fact that the reduction of
%quasiparticle weight (self-energy effect) due to electron correlation
%plays a negative role in enhancing the pairing susceptibility.

%\bibliography{FeAspc2,qmc,myownpaper}

\begin{thebibliography}{43}
\bibitem[{\citenamefont{t1}(2012)}]{t1}
\bibinfo{journal}{The signs of $t_1$ being the same in the $x$-and $y$-direction suggests the
orbital character of the model can not be pure $d_{xz}$ and $d_{yz}$
orbitals. The orbitals in the $S_4$ model are heavily dressed by the $d_{xy}$
orbital, and the Wannier functions for the two orbitals of the $S_4$ model shall be reported in our further work.}
\bibitem[{\citenamefont{pairing}(2012)}]{pairing}
\bibinfo{journal}{This is not obvious in the present paper since only one $S_4$ isospoin is considered. However, one can draw this conclusion by considering the parity requirement for spin signlet pairing.}
\end{thebibliography}

\end{document}